\documentclass{aastex631}
\usepackage{natbib}
\usepackage{verbatim}

\usepackage{amssymb, amsmath}

\begin{document}

\title{Comments on the Union3 ``Spline-Interpolated Distance Moduli'' Model}
\author[0000-0001-6315-8743]{Alex G.\ Kim}
\affiliation{Lawrence Berkeley National Laboratory, Berkeley, CA 94720 USA}
\email{agkim@lbl.gov}

\begin{abstract}
The Union3 ``Spline-Interpolated Distance Moduli'' model posterior has been distributed for third-party cosmology analysis. The posterior prefers a large value of
$\Omega_M$, a small absolute value of $w_0$, and a negative $w_a$, but still accommodates  $\Lambda$CDM; the supernova data alone are not strongly constraining.  The posterior is built assuming an underlying model and prior, both of which must be made to conform with any new model and prior being analyzed.
The posterior is calculated for
a  prior that is not flat but rather has non-trivial structure in $\Omega_M$--$w_0$--$w_a$;
the associated likelihood is slightly shifted relative to the posterior.
The  posterior for a prior that is flat in $\Omega_M$--$w_0$--$w_a$ is also shifted relative to the original, but not
at a level that is  statistically significant.    The misapplication of Union3 results in the  ``DESI2024 VI'' cosmology
fits are inconsequential.
\end{abstract}

\section{Introduction}
Supernova, BAO, and CMB data are used to fit cosmological and dark-energy parameters in the DESI cosmology
analysis in the paper ``DESI2024 VI''  \citep{2024arXiv240403002D}.  The fit to the flat $w_0$--$w_a$ model is discrepant with the $\Lambda$CDM
model at the 3.5$\sigma$ level for the combination of DESI+CMB with Union3  \cite{2023arXiv231112098R}.
Similar level discrepancies are found for the Pantheon+ \cite{2022ApJ...938..113S} and the DES-SN5YR\footnote{Data available at \url{https://github.com/des-science/DES-SN5YR.}} supernova compilations.

The Union3 ``Spline-Interpolated Distance Moduli'' model is designed to compress the  supernova data for convenient use by the community.\footnote{Union3 data and code are available at \url{https://github.com/rubind/union3_release}.}
The  model distance modulus
is the sum of the distance modulus of $\Omega_M=0.3$  $\Lambda$CDM plus a second-order spline specified by node values ($n$)
at a fixed set of 22 redshifts. 
The parameters $n$ are equivalent to the distance moduli for a fixed set of redshifts.
Standard normal distributions are used as priors for $n$.  
The posterior of the nodes is approximated as a Gaussian, and released as the location and Hessian at maximum.
In the Union3 paper, cosmological inference is not made from the Spline model but rather directly from physics-based models.

The Spline model has much more flexibility than the physics-based models of interest, e.g.\ those involving, $\Omega_M$,
$w_0$, $w_a$.  Though not precisely correct, those physics-based models can be considered as being embedded in the Spline
model such that the Union3  posterior can  be used as a prior or, divided by the Union3 prior, as a likelihood in joint-probe analyses.   It is worth reviewing
how the Union3 results should be incorporated into a joint analysis; indeed the ``DESI2024 VI''
cosmology paper  \cite{2024arXiv240403002D} mixes inconsistent priors in its analysis.

The objectives of this note are to: Explore what the Spline-model posterior says about $w_0w_a\Lambda$CDM cosmology, which is not normally
fit with  supernova data alone (\S\ref{sec:union3}); Transform the Union3 prior in $n$ to $\Omega_M$--$w_0$--$w_a$
(\S\ref{sec:prior});
Examine the new posterior for the DESI prior
(\S\ref{sec:flatposterior}); Show how to include the Union3 Spline posterior in a joint analysis of a different model (\S\ref{sec:joint}).
The note ends with a few conclusions (\S\ref{sec:conclusions}).

\section{Union3 Spline Model and Posterior}
\label{sec:union3}
Union3 considers several cosmological models, including  flat and open $\Lambda$CDM, $w$CDM, and $w_0w_a$CDM,
with and without non-supernova data.
In addition to these standard models,  Union3 analyzes supernova data only using a Spline model where the  distance modulus
is the sum of the distance modulus of $\Omega_M=0.3$  $\Lambda$CDM plus a second-order spline specified by node values, $n$,
at a fixed set of 22 redshifts: $\mu(z;n) = \mu_{\Lambda \text{CDM}}(z;\Omega_M=0.3) + \text{spline}(z;n)$.  The distance modulus of this model has significantly more flexibility 
and hence retains more information about the underlying data compared to the physics-motivated models.
The model parameters correspond to the value of distance modulus at 22 redshifts.
Union3 uses Bayesian statistics for parameter estimation: as such it requires 
priors on the node values, which are taken to be standard normal distributions
$p^\text{Union3}_\text{prior}(n)=  \mathcal{N}(n,1)$.
The posterior of this model fit is being made public and used by the community.

The physics cosmologies (e.g.\ $w_0w_a$CDM)  are not embedded in the Spline model but 
they come close; any cosmology's distance moduli at the node redshifts can be exactly replicated by the Spline model, acknowledging that
there are
inconsistencies between the nodes.  Neglecting this, lets say that the flat $w_0w_a$CDM model
is a subspace of the Spline model defined by the mapping
\begin{align}
	n &= f(\Omega_M, w_0, w_a; z) \\
	& \equiv \mu_{w_0 w_a \text{CDM}}(z;\Omega_M, w_0, w_a)  - \mu_{\Lambda \text{CDM}}(z;\Omega_M=0.3). \label{eqn:f}
\end{align}

\begin{figure}[htbp] 
   \centering
   \includegraphics[width=5.5in]{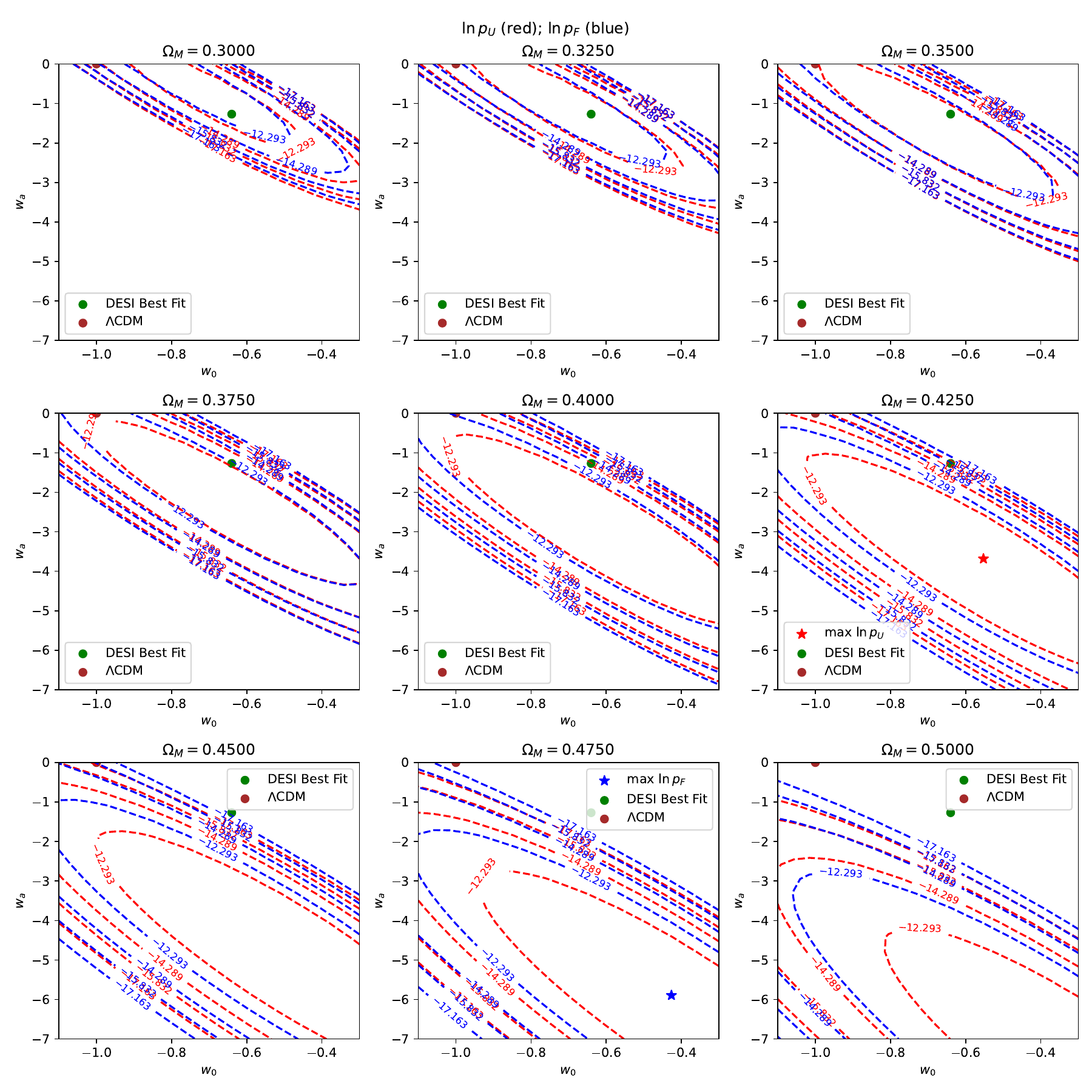} 
   \caption{Contours of two different Union3 posteriors for the Spline-model parameters for a grid of values
 $w_0$--$w_a$ and $\Omega_M$.   
  Red: For the fiducial prior.  Contours of  $\ln{p_U}+c$, where $c$ is chosen such that $\ln{p_U}+c$  is equivalent to  $-\chi^2/2$. Contour levels are set to 68.27\%, 84.28\%, 91.67\%, and 95.55\% confidence intervals.
   Blue: For the DESI prior, which is flat in the plotted range for these parameters. 
   The blue contour levels of $\ln{p_F}+c$ are the same as those for $p_U$ but do not correspond to the same confidence intervals.
   The maximum of the $\ln{p}_U$  ($\ln{p}_F)$ posterior in this space is shown as the red (blue) star.  (The absolute maximum
   of $p_U$
   lies outside the $\Omega_M$--$w_0$--$w_a$ space.)
   Points show the position of DESI's  BAO+CMB+Union3 flat $w_0w_a$ best-fit 
    and  $\Lambda$CDM.}
   \label{fig:posterior}
\end{figure}

While the Union3 posterior (denoted as $p_U$) is for the 22-dimensional node space, it is of interest to examine the posterior values 
as a function of $\Omega_M$--$w_0$--$w_a$ for the
subspace where Eq.~\ref{eqn:f} holds. 
Values of (the Gaussian approximation of) $\ln{p_U}$
 are presented as the red contours in Figure~\ref{fig:posterior}, where the levels are set to 68.27\%, 84.28\%, 91.67\%, and 95.55\% confidence intervals..
The largest $\ln{p_U}$ in $w_0w_a$CDM is at $\Omega_M\approx 0.425$, $w_0 \approx -0.55$, and $w_a \approx -3.7$. $\Lambda$CDM at $\Omega_M \approx 0.35$ is within the 68\% confidence region.
Supernova data alone prefer high $\Omega_M$, a low absolute value of $w_0$, and a negative $w_a$, but do not have
the constraining power to exclude standard cosmology.

The maximum $\chi^2/2$ in the plotted space is $10.33$.  While $w_0w_a\Lambda$CDM is allowed, the best fit ($\chi^2=0$)
is at a value of $n$ that is not represented by that model. 

\section{Union3 Prior for $n$ in Terms of $\Omega_M$--$w_0$--$w_a$}
\label{sec:prior}
The Union3 prior for the Spline analysis is designed to keep distance moduli close to a reasonable
fiducial distance modulus function.  It is of interest to see what this prior, which is defined in magnitude
space, looks like in terms of cosmological and dark-energy parameters.

The standard normal prior on the nodes $\mathcal{N}(a,1)$ corresponds to a prior in  $\Omega_M$--$w_0$--$w_a$ of
\begin{align}
p^\text{Union3}_\text{prior}(\Omega_M, w_0,w_a)  & =p^\text{Union3}_\text{prior}(n=f(\Omega_M, w_0, w_a; z))  \sqrt{\det{\left(J^T J\right)}} \\
& = \mathcal{N}(f(\Omega_M, w_0, w_a; z),1)  \sqrt{\det{\left(J^T J\right)}}, \label{eq:Union3prior}
\end{align}
where $J$ is the Jacobian matrix of $f$ and $J^TJ$ is the Gram matrix.

\begin{figure}[htbp] 
   \centering
   \includegraphics[width=5.5in]{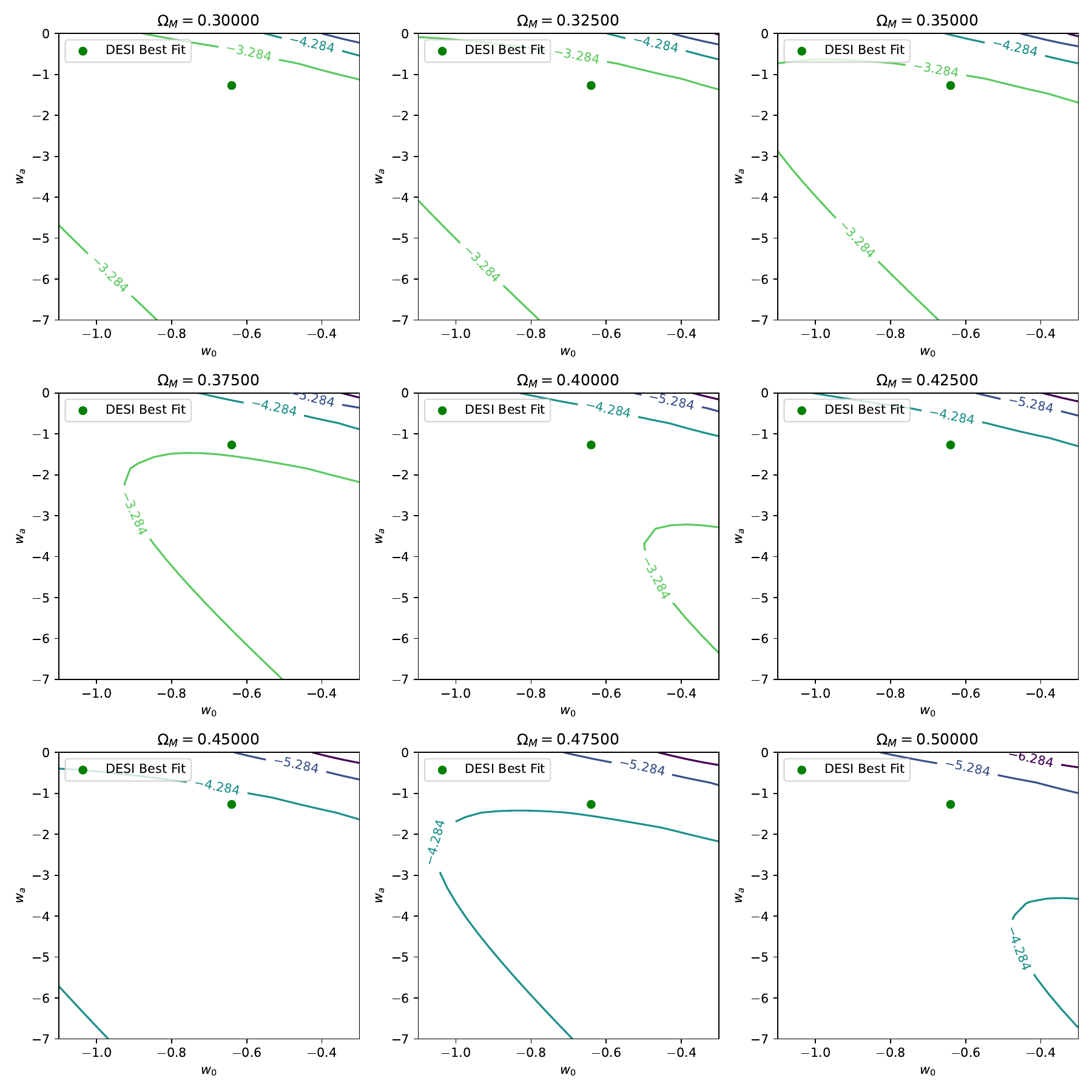} 
   \caption{Surfaces of the Union3 prior, originally specified in terms of node parameters, transformed to the $w_0w_a$CDM parameterization
   $\ln{p^\text{Union3}_\text{prior}}(\Omega_M, w_0,w_a)$,  for a grid of values
 $w_0$--$w_a$ and $\Omega_M$.   
   Points show the position of the BAO+CMB+Union3 flat $w_0w_a$ best-fit.}
   \label{fig:priors}
\end{figure}

Surfaces of  $\ln{p^\text{Union3}_\text{prior}}(\Omega_M, w_0,w_a)$ values\ for a grid of $\Omega_M$--$w_0$--$w_a$ are shown in
Figure~\ref{fig:priors}.   The prior is not uniform in  $\Omega_M$--$w_0$--$w_a$.
The prior is not strongly informative, the contours of the $p_U$
posterior shown in Figure~\ref{fig:posterior} do not have the same shape of the prior contours.  Nevertheless,
we will see in the  Section~\ref{sec:flatposterior} that a different reasonable prior can shift the posterior
by an appreciable amount.

\section{Posterior for a Flat  $\Omega_M$--$w_0$--$w_a$  Prior}
\label{sec:flatposterior}
The Union3  posterior for a different choice of prior can be obtained through
\begin{align}
p(n) & = p_U(n)\frac{p_\text{prior}(n)}{p^\text{Union3}_\text{prior}(n)}\\
 & = p_U(n) \frac{1}{p^\text{Union3}_\text{prior}(n) } \frac{p_\text{prior}(\theta)}{ \sqrt{\det{\left(J^T J\right)}}} ,
\end{align}
where the prior is given in terms of parameters $\theta$
related to the node values as $n=f(\theta)$ and $J$ is the Jacobian of $f$.
For example, 
the posterior for the 
DESI prior $\Omega_M \in \mathcal{U}(0.01,0.99)$, $w_0 \in \mathcal{U}(-3,1)$, $w_a \in \mathcal{U}(-3,2)$
is
\begin{align}
p_F(n) 
& = \frac{p_U(n)}{(0.98)(4)(5) \mathcal{N}(n,1)  \sqrt{\det{\left(J^T J\right)}}}.
\label{eq:flatprior}
\end{align}
The above equation only applies over the range of $n$ allowed by  $w_0w_a$CDM, the posterior is otherwise zero.

Values of $\ln{p}_F$ are shown for  $\Omega_M$--$w_0$--$w_a$ as blue contours in Figure~\ref{fig:posterior}.
The largest $\ln{p_F}$ in $w_0w_a$CDM is a $\Omega_M\approx 0.475$, $w_0 \approx -0.425$, and $w_a \approx -5.9$.  This is a shift from the original Union3 posterior, though  the $\ln{p_F}$ maximum is well within
the $\ln{p_U}$ posterior's 68\% confidence region.

\section{Union3 in Joint Analyses of ``Embedded'' Models}
\label{sec:joint}
Union3 results can be combined with other datasets in multi-probe analyses of models that are embedded in the Spline model.
Consider independent data $d$ used to fit a model parameterized by $\theta$
that predicts distance moduli at the Spline-model redshifts as $n=f(\theta)$.
The Union3 posterior can be used as a prior for $n$, or equivalently, divided by the prior to
give the Union3 likelihood.  The posterior of the joint analysis using Union3 as a prior is
\begin{align}
p(\theta|d) &\propto \int  \mathcal{L}(\theta; d)p(\theta|n)p_U(n) \frac{p_\text{prior}(n)}{p^\text{Union3}_\text{prior}(n)}dn\\
&=  \mathcal{L}(\theta; d)    {\sqrt{\det{J^TJ}}}   {p_U(f(\theta))}  \frac{p_\text{prior}(f(\theta))}{p^\text{Union3}_\text{prior}(f(\theta))}\\
&=  \mathcal{L}(\theta; d)  {\sqrt{\det{J^TJ}}}  {p_U(f(\theta))}  \frac{p_\text{prior}(\theta)}{  {\sqrt{\det{J^TJ}}}  p^\text{Union3}_\text{prior}(f(\theta))}, \label{eq:correct} \\
&=  \mathcal{L}(\theta; d) {p_U(f(\theta))}  \frac{p_\text{prior}(\theta)}{ p^\text{Union3}_\text{prior}(f(\theta))} \\
&\propto  \mathcal{L}(\theta; d) \mathcal{L}_\text{Union3}(f(\theta)) {p_\text{prior}(\theta)}, \label{eq:correct}
\end{align}
 where  $J$ is the Jacobian of $f$,
$\mathcal{L}(\theta; d)$ is the likelihood of the complementary dataset, and
\begin{equation}
\mathcal{L}_\text{Union3} \propto \frac{p_U}{ p^\text{Union3}_\text{prior}}
 \end{equation}
is the Union3 likelihood.

\begin{figure}[htbp] 
   \centering
   \includegraphics[width=5.5in]{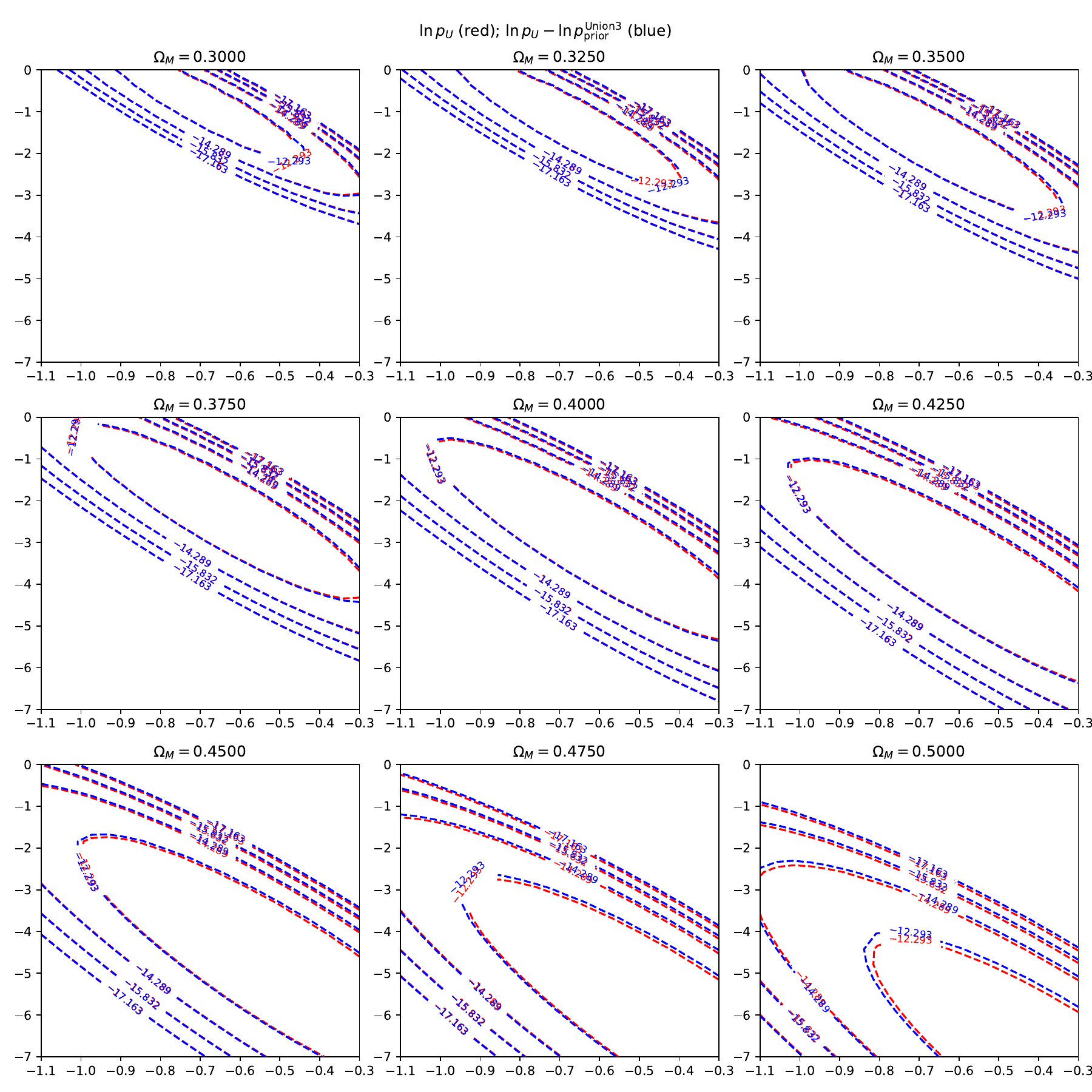} 
   \caption{Surfaces of the Union3 posterior $\ln{p_U}$ (red) and likelihood $\ln{\mathcal{L}_\text{Union3}}  = \ln{p_U}-\ln{p_\text{prior}^\text{Union3}}$ (blue).   The red curves
   and choice of contours levels are the same as in Fig.~\ref{fig:posterior}.\label{fig:likelihood}}
\end{figure}

The DESI paper analysis uses
\begin{align}
p(\theta | d) & \propto  \mathcal{L}(f(\theta);d) {p_U(f(\theta))}  p^\text{DESI}_\text{prior}(\theta) \label{eq:DESI}.
\end{align}
The above posterior does not remove the Union3 prior: when the DESI analysis
was being performed an early draft of the Union3 paper mistakenly reported its priors as being flat.    The Union3 posterior and likelihood are plotted
in Figure~\ref{fig:likelihood}; clearly this omission does not strongly alter the conclusions in the ``DESI2024 VI''
paper. 

DESI also approximates the Union3 posterior by a Gaussian ${p_U(f(\theta))} \approx \mathcal{N}( \hat{n}-f(\theta) , C_U)$,
where $\hat{n}$ and $C_U$ were distributed by the Union3 team.
Caution should be exercised when multi-probe analyses favor regions far from the Union3 best fit  where the Hessian approximation of the
posterior may be less secure.

Supernova data are not expected to be Gaussian distributed around a mean with known variance, so it is not possible
to make frequentist inferences from the likelihood alone.  The statistic
$-2\ln{\mathcal{L}_\text{Union3}}$  is not drawn from a $\chi^2$-distribution and its $p$-values would have to be determined
independently.

\section{Conclusions}
\label{sec:conclusions}
Multi-probe  analysis is required to get the most cosmological information out of the plethora of available data. 
Compressed data products from different experimental groups can be combined to achieve this purpose.
Although not as optimal as simultaneously fitting all data, this approach simplifies algorithms and allows a larger
fraction of
the community to work with the data.  Distributing calculated likelihoods is simpler than distributing data and expecting others to implement the likelihood algorithm.

As such, Union3 has released a posterior for a general model that summarizes the Hubble diagram inferred from its data.
The model is parameterized by distance moduli on a grid of redshifts and is (approximately) a superspace that
contains a broad range of models.    I have compared the shapes of the Union3 likelihood and posteriors for two
reasonable priors and find that differences are not significant.

Likelihoods and posteriors are model-dependent.  It is important to keep in mind that any new model may not perfectly represented by the 
distribution model.

To enable more accurate work, Union3 will provide a higher-order description of its posterior and full set of MCMC chains.

\begin{acknowledgments}
I thank Kushal Lodha, William Matthewson, and Arman Shafieloo for discussions that inspired this work, 
and Patrick McDonald for his invaluable insight into the DESI analysis.
This work was supported by the U.S. Department of Energy (DOE), Office
of Science, Office of High-Energy Physics, under Contract No. DE–AC02–05CH11231.
\end{acknowledgments}

\bibliographystyle{aasjournal}
\bibliography{apj-jour,ref}

\end{document}